\documentclass[11pt,twoside]{article}
\usepackage{asp2010}

\resetcounters

\begin{document}

\title{On the M31 Nova Progenitor Population}
\author{S.~C. Williams$^1$, M.~J. Darnley$^1$, M.~F. Bode$^1$, and A.~W. Shafter$^2$
\affil{$^1$Astrophysics Research Institute, Liverpool John Moores University, Twelve Quays House, Egerton Wharf, Birkenhead, CH41~1LD, UK}
\affil{$^2$Department of Astronomy, San Diego State University, San Diego, CA 92182, USA}}

\begin{abstract}
We present a survey of M31 novae in quiescence.  This is the first catalogue of extragalactic systems in quiescence and contains 37 spectroscopically confirmed novae from 2006 to 2013.  We used Liverpool Telescope and Faulkes Telescope North images taken during outburst to identify accurate positions for each system.  These positions were then transformed to archival {\it Hubble Space Telescope (HST)} images and we performed photometry on any resolvable source that was consistent with the transformed positions.  As red giants in M31 will be resolvable in the {\it HST} images, we can detect systems with red giant secondaries.  There are only a few confirmed examples of such systems in our Galaxy (e.g.\ RS~Oph and T~CrB).  However, we find a much higher portion of the nova population in M31 may contain red giant secondaries.  For some novae, coincident {\it HST} images had been taken when the nova was still fading, allowing us to produce light curves that go fainter than is possible to achieve for most extragalactic systems.  Finally, we compare the M31 and Galactic quiescent nova populations.
\end{abstract}

\section{Introduction}

Amongst the hundreds of known Galactic classical novae (CNe), there are only ten confirmed recurrent novae \citep[RNe;][]{2010ApJS..187..275S}.  Traditionally, RNe have only been confirmed by the observation of more than one outburst from the same system.  Using multiple outbursts as the sole discriminator between recurrent and classical novae is subject to a number of significant selection effects, as such the Galactic population of ten RNe is a lower limit and quite likely far from the true picture.  Recently, more indirect methods have been employed to introduce a small number of new candidate RNe in systems with only one recorded outburst (see paper by M.~J.~Darnley, these proceedings).

The Sun's position in the Milky Way significantly hinders any attempt to study the nova population of the Galaxy.  As such, traditionally, we turn to M31 which, whilst far from ideal, provides us with the best opportunity to study the nova population of an entire galaxy.  With a nova rate of $65^{+16}_{-15}\;\mathrm{yr}^{-1}$ \citep{2006MNRAS.369..257D}, M31 presents us with a potential sample size twice that of the Milky Way and far beyond the $\sim10$ Galactic novae that are practically observable each year.

Ones ability to detect RNe in M31 is still hampered by our inability to observe M31 for half of each year; large gaps between the observations and surveys of M31; and poor archival astrometry \citep[see, for example][]{2009ApJ...705.1056B}, amongst many other effects.  Nonetheless, there have been a number of attempts to explore the RN population of M31 by searching for multiple outbursts, but these have only yielded a handful of RN candidates (see paper by A.~W.~Shafter, these proceedings).  This approach is further hampered by the misidentification of long period Mira variables as M31 novae \citep[see, for example][]{2004MNRAS.353..571D}.

When studying M31 novae, a number of the above problems can be overcome by using a spectroscopically confirmed sample, with well determined astrometry, but this essentially limits us to M31 novae since around 2006 \citep{2011ApJ...734...12S}.  Based on the recurrence timescales of their Galactic counterparts, such a short baseline is not long enough to recover a good sample of RNe using multiple outbursts alone.

Here we propose a technique that can be used to recover the progenitor systems of novae containing red giant secondaries - those belonging to the RG-nova class \cite[which is dominated by confirmed and candidate RNe of the RS~Oph sub-class,][]{2012ApJ...746...61D}.  In some (exceptional) cases this technique may also be able to recover novae with sub-giant secondaries (SG-novae; a class dominated by U~Sco-like RNe) as was recently achieved for the (albeit significantly closer) LMC recurrent nova LMC~2009a (Bode et al., in prep).

\section{Observations}

A number of space-based and the larger ground-based optical telescopes are capable of resolving the red giant population of M31.  As such, the progenitor systems of RG-novae can in principle be directly imaged.  This was successfully carried out for the RN candidate M31N~2007-12b \citep[see][on which this work is largely based]{2009ApJ...705.1056B}.  This technique relies upon accurate registration between images of the nova in outburst and deeper (typically archival) high spatial resolution images when the system is in quiescence (either post- or pre-outburst).  For this study, we use data taken by the Liverpool Telescope (LT) and one of its sister telescopes, Faulkes Telescope North (FTN), to determine the outburst position of the novae and archival {\it Hubble Space Telescope (HST)} data for the identification and photometry of the progenitor.  The {\it HST} observations are taken from a mixture of the WFPC2, ACS/WFC and WFC3/UVIS instruments, all of which provide a very good overlap with the LT/FTN fields.  A full description of the progenitor recovery technique is given in Williams \& Darnley et~al.\ (in prep) and is summarised in \citet{2009ApJ...705.1056B}.

\section{Progenitor Results}

\begin{figure}
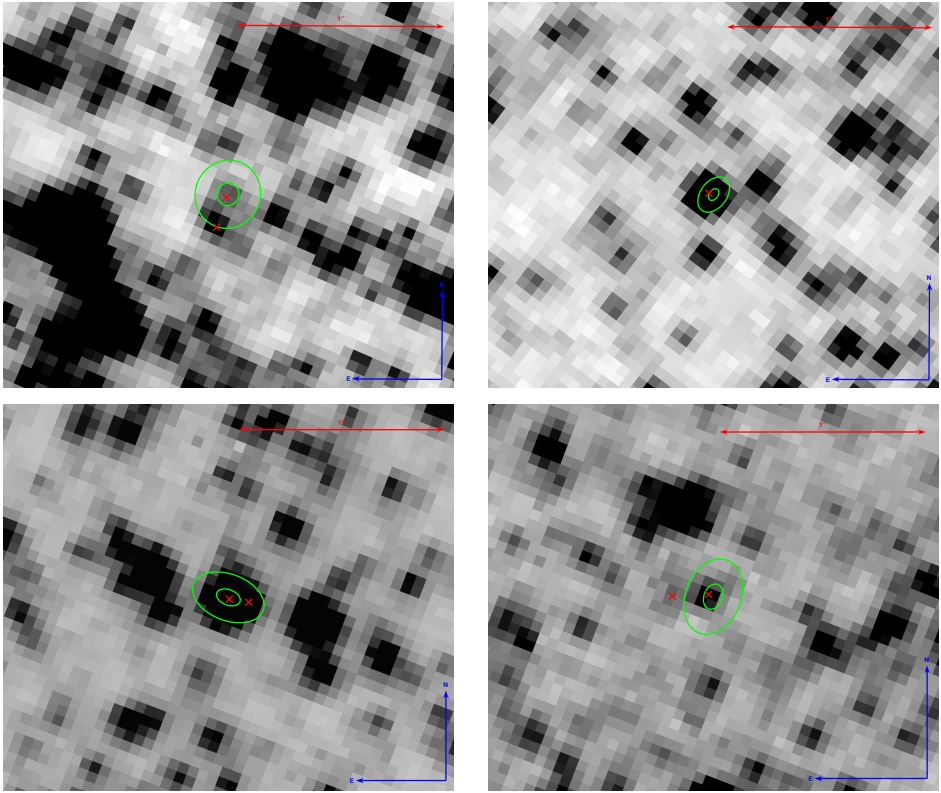

\plotfour{2007-02b}{2007-10a}{2007-11d}{2009-11d}
\caption{{\it HST} ACS/WFC images of the $\sim3^{\prime\prime}\times3^{\prime\prime}$ region surrounding the novae (from top-left, clockwise) M31N~2007-02b, 2007-10a, 2009-11d and 2007-11d.  The inner green circle indicates the $1\sigma$ radius search region for the progenitor, the outer circle the $3\sigma$ region, and the red $\times$ indicates the position of progenitor candidates.\label{HST}}
\end{figure}

From a spectroscopically confirmed sample of 108 M31 nova from 2006 onward \citep[most from][]{2011ApJ...734...12S}, 37 have accurate outburst astrometry from LT/FTN data coupled with quiescent archival {\it HST} imaging.  Initial analysis shows that of these 37, the position of at least 23 novae are coincident (at the $3\sigma$ level, or better) with at least one resolved source in the archival data.  Of this subset, a Monte-Carlo analysis indicates that 10 of these alignments would be expected to occur through chance less than 5\% of the time.  Figure~\ref{HST} presents thumbnail {\it HST} images centred around four of the better progenitor candidates; M31N~2007-02b, 2007-10a, 2007-11d and 2009-11d all of which have alignments $\le1\sigma$ and coincidence probabilities $<3\%$.

\begin{figure}
\plotone{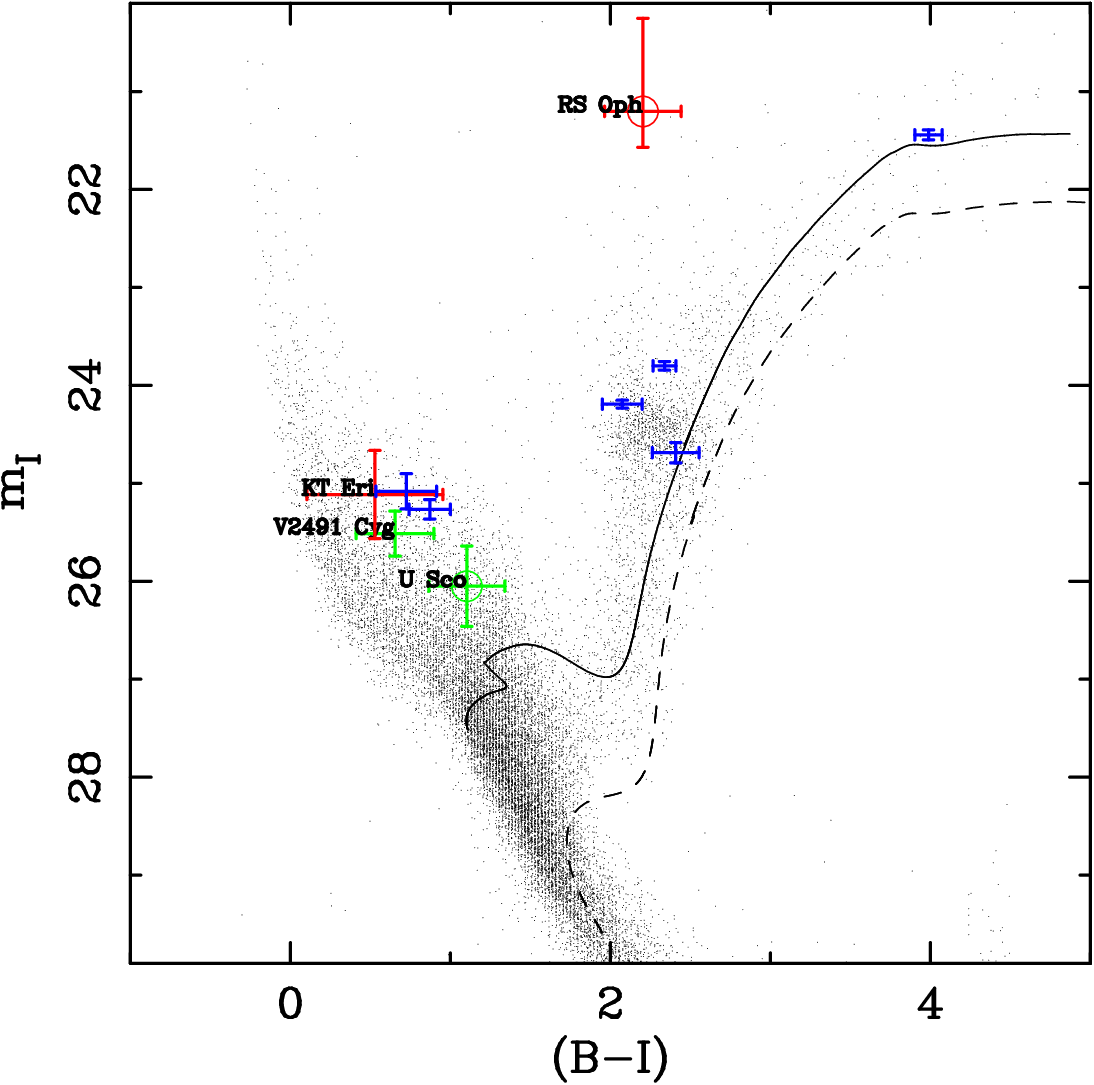}
\caption{Colour-magnitude diagram showing stars from the {\it Hipparcos} data set \citep{1997ESASP1200.....P} moved to the position of M31 assuming $(m-M)_{0}=24.3$ \citep{1990ApJ...365..186F} and estimated extinction of $E_{B-V}=0.1$~mag \citep[no attempt has been made here to correct for internal extinction]{1992ApJS...79...77S}.  Green and red points show Galactic members or candidates of the SG-Nova class and RG-Nova class, respectively (as they would appear within M31), and the blue points show photometry of six candidate M31 nova progenitor systems. The known recurrent novae in this sample have been identiﬁed by an additional circle. The black dashed line shows the evolutionary track of a 1~M$_{\odot}$ solar-like star, the solid line a 1.4~M$_{\odot}$ solar-like star \citep{2004ApJ...612..168P}.\label{progen}}
\end{figure}

Figure~\ref{progen} shows an ``M31'' colour-magnitude diagram and presents the photometry of six M31 nova progenitor candidates.  Four of these systems are associated with the red giant branch and would likely be classified as RG-novae.  However, two of these systems are coincident with the quiescent positions of the SG-novae U~Sco and V2491~Cyg, which implies that they may also be SG-novae.  It is worth noting that the RG-nova candidate KT~Eri has particularly large $B$-band emission and is also coincident with this pair of systems.

The observed Galactic RG-nova population is $\sim2-3\%$ that of the Galactic nova population \citep{2012ApJ...746...61D}, our initial results for M31 indicate that as many as 10 from a sample of 37 spectroscopically confimed novae are associated with a red giant secondary.  This indicates at first sight a M31 RG-nova population of $\sim30\%$ the underlying nova population, significantly higher than has been observed Galactically.

\section{Extended Light-Curve Coverage}

\begin{figure}
\plotone{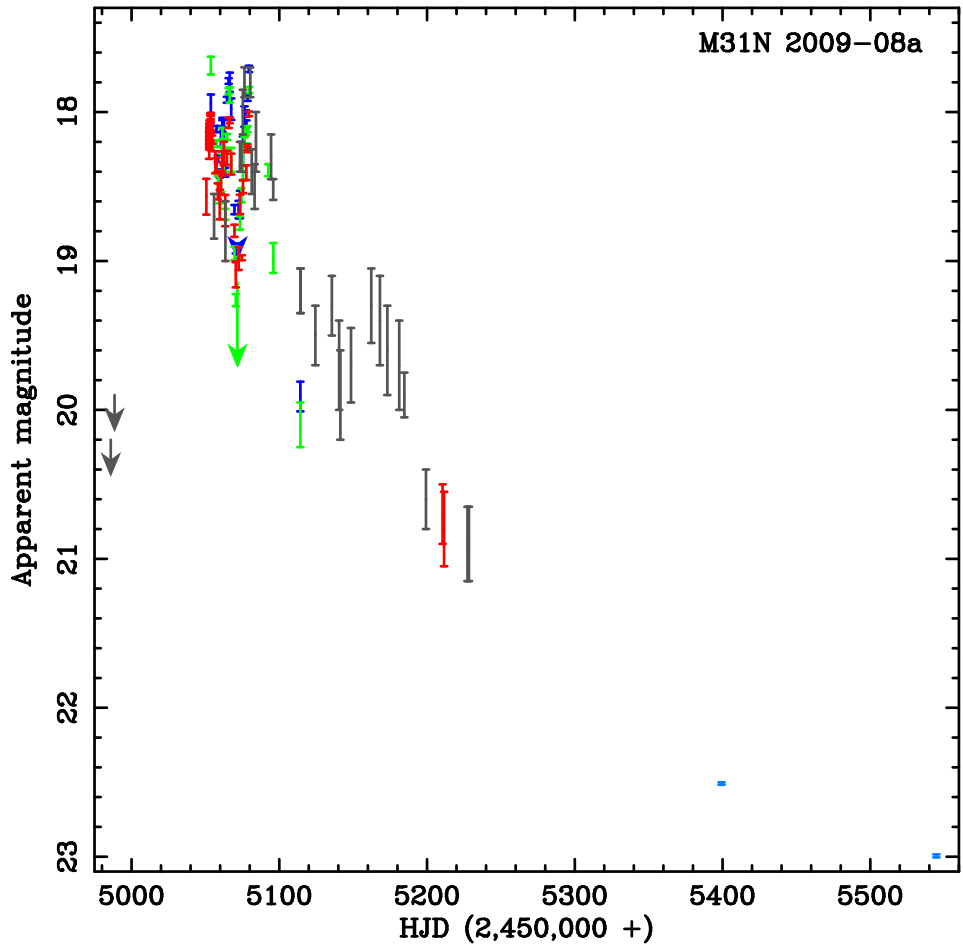}
\caption{Light curve of M31N~2009-08a, ground-based data from \citet{2011ApJ...734...12S}.  The uncertainties in the photometric measurements are shown as vertical bars with the following colours representing the different band passes: $B$, blue; $V$, green; $R$, dark grey; $r'$, red; $i'$, black; $z'$, light gray; {\it HST} ACS/WFC F475W, light blue. Upper flux limits are indicated by downward facing arrows.\label{lc}}
\end{figure}

In a small number of chance cases, archival {\it HST} data have been taken shortly following the outburst of M31 novae.  As such this has allowed us to follow the evolution of M31 novae to previously unprecedented deepness.  In Figure~\ref{lc} we present the light curve of M31N~2009-08a \citep[originally presented by][]{2011ApJ...734...12S}.  This very slow nova was originally followed by a combination of telescopes (including LT/FTN) through $\sim3$ magnitudes below peak.  Additional {\it HST} ACS/WFC archival data have allowed us to follow this nova to $\sim5$ magnitudes below peak, at a time almost 500~days post outburst.

\bibliographystyle{asp2010}
\bibliography{williams}

\end{document}